\documentclass[twocolumn,twoside]{revtex4-2}
\usepackage{graphicx}
\usepackage{inputenc}
\usepackage{subfig}
\usepackage{color}
\usepackage{hyperref}
\usepackage{amsmath}

\newcommand\siesta{\textsc{Siesta}}

\let\etal\relax
\newcommand{\etal}{\mbox{\textit{et al.}}~}

\newcommand{\Eqref}[1]{Eq.~(\ref{#1})}
\newcommand{\Figref}[1]{Fig.~\ref{#1}}

\newcommand{\eg}{{\em e.\,g. }}
\newcommand{\ie}{{\em i.\,e. }}

\newcommand{\bfk}{\mathbf{k}}
\newcommand{\bfR}{\mathbf{R}}
\newcommand{\bfv}{\mathbf{v}}
\newcommand{\Vdir}{-\hat{\mathbf{e}}}
\begin{document}

\title{Simple approach to current-induced effects -- bond weakening in metal chains}

\author{Nick Papior}
\affiliation{Department of Applied Mathematics and Computer Science, Technical University
    of Denmark, DK-2800 Kongens Lyngby, Denmark}

\author{Susanne Leitherer}
\affiliation{Department of Chemistry, University of Copenhagen, DK-2100 Copenhagen, Denmark}
\author{Mads Brandbyge}
\affiliation{Center for Nanostructured Graphene}
\affiliation{Department of Physics, Technical University of Denmark, DK-2800 Kongens Lyngby, Denmark}


%

\date{\today}

\begin{abstract}
We present a simple, first principles scheme for calculating mechanical properties of nonequilibrium bulk systems assuming an ideal ballistic distribution function for the electronic states described by the external voltage bias. This allows for fast calculations of estimates of the current-induced stresses inside bulk systems carrying a ballistic current. The stress is calculated using the Hellmann-Feynman theorem, and is in agreement with the derivative of the nonequilibrium free energy. We illustrate the theory and present results for one-dimensional (1D) metal chains.  We find that the susceptibility of the yield stress to the applied voltage agrees with the ordering of break voltages among the metals found in experiments. 
In particular, gold is seen to be the most stable under strong current, while aluminum is the least stable. 
\end{abstract}


\maketitle

Metallic interconnects and their stability under strong electrical current, \ie electronic nonequilibrium conditions, plays a central role in the on-going down-scaling of electronic devices\cite{HoffmannVogel2017}. The ultimate limit of passing current through contacts containing a few and down to a single atom in the cross section has been investigated for more than three decades\cite{Gimzewski1987, Takayanagi98, Agrait2003}. For conductors in the atomic limit the electrons essentially move through the contact without loosing energy to atomic vibrations.
Therefore, single atom wide contacts and chains of a range of common metals\cite{YaSa.1997,Mizobata2003,Smit2004,Wakasugi2017} can sustain voltages on the order of 1 V, which corresponds to extreme current densities on the order of $10^{10}\,\mathrm{A/cm}^2$.
The contact disruption taking place at high voltage and current is still not well understood. Different mechanisms have been put forth in order to understand the role of nonequilibrium for the stability of the atomic contacts. Joule heating in the contacts\cite{Todorov1998,Smit2004}, as well as the effect of the electric field are suggested to be important factors, along with the role of heat-transfer between contact and bulk electrodes\cite{Engelund2009}. Furthermore, the action of the current-induced/nonequilibrium ``wind'' forces\cite{DuMcTo.2009,Lu2010}, which may transfer energy to the vibrations beyond the Joule heating effects, may even lead to structural instabilities (``runaway'') behavior at particular critical voltages on the order of 1V. This effect might explain the different breaking modes found for longer atomic chains of Au\cite{Sabater2015}. 
Clearly, vibrational excitation along with the ambient temperature, heat conductivity\cite{Cui2017}, energy barriers related to the bond-breaking, and detailed atomic structure\cite{Solomon17} are important factors in this complicated process.

Despite the complications, the experiments, typically involving large statistical samples, show a rather clear distinction between the current-induced disruption or switching behavior of atomic contacts of different metals. For instance, short atomic Au chains formed at low temperature were shown to break at voltages around 1-2V, while it was already around $0.5$V for Pt \cite{Sabater2015}.
Very recently, Ring \etal{} \cite{Ring2020} showed in comparative studies how switching occurred at decreasing voltages in the sequence Au, Cu, Pb, Al for atomic contacts with conductances up to 6 $G_0$ ($G_0=2e^2/h$). It was noted that this sequence did not correspond to the sequence in melting or Debye temperatures. 
Furthermore, extensive first principles molecular dynamics calculations including the coupling of current to phonons\cite{Ring2020} (Joule and wind-force) described by density functional theory (DFT), yielded a magnitude of break voltages in agreement with the experiments. However, these calculations neglected the nonequilibrium change in bond strength, and, notably, were not able to reproduce the material stability sequence and found that Al was highest and Cu lowest in switching voltage.

Earlier calculations have demonstrated an ``imbrittlement''/weakening of metallic bonds in the presence of current\cite{Todorov2001}. This was related to the nonequilibrium charge redistribution and a decrease in the bonding-charge residing between the atoms or overlap population, as calculated by density functional theory combined with  nonequilibrium Greens function methods (DFT-NEGF)\cite{Brandbyge2003}. More recently, the change in bonding forces in a C$_{60}$-C$_{60}$ contact carrying a current has been measured, and was explained in terms of this type of nonequilibrium charge-redistribution in the system\cite{Brand2019}.

In this paper we introduce a conceptually simple, approximate method based on density functional theory with standard periodic boundary conditions which enables us to calculate the bond-weakening/``imbrittlement'' in the presence of the nonequilibrium charge redistribution due to current alone, neglecting effects of scattering-dipole fields\cite{Landauer1957}. We employ it to assess how the yield strength of single-atom metal chains changes with applied bias and relate the bond-weakening to the underlying electronic structure. Interestingly, we find that the bond-weakening with bias follows the material sequence (Au, Cu, Pb, Al) seen in the recent comparative experiments\cite{Ring2020}.

\begin{figure}[tbh]
	\includegraphics[width=0.9\linewidth]{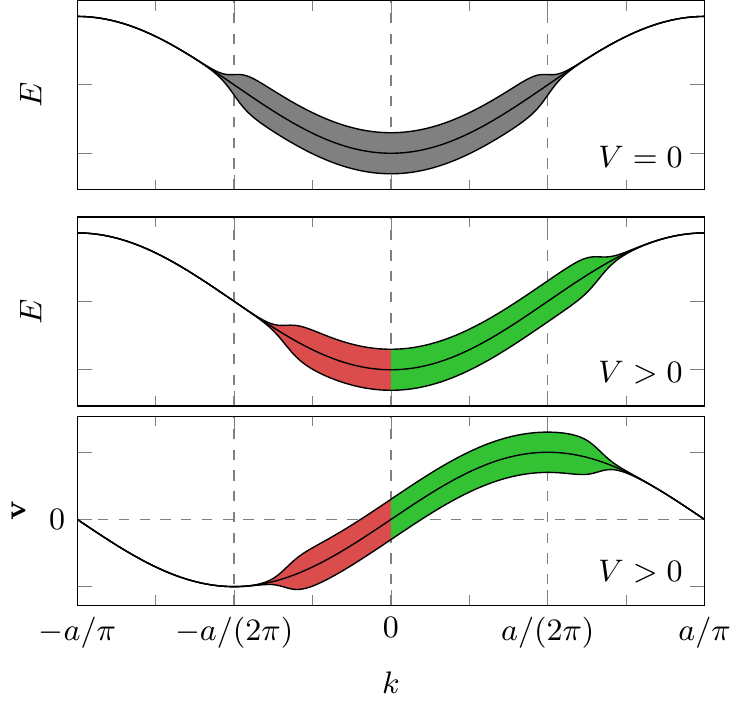}
	\caption{Example of bulk-bias applied to a model 1D chain system. Top/Middle: Band and occupation (filling) of states at zero/finite bias. Bottom: Band velocity and occupation. The filled area corresponds to the Fermi-Dirac distribution of the band and red/green parts are ``left''/``right'' moving states. This distribution equates to a shift of the local chemical potential according to the direction of the electron w.r.t. the applied bias direction.}
	\label{fig:1Dexample}
\end{figure}

\section{Method}
\label{sec:bulk}

The concept of the Landauer resistivity-dipole\cite{Landauer1957} yielding a local potential drop around the region where electrons are scattered is well established and observed in experiments\cite{Homoth2009}. For defect-free, one-dimensional conductors connected to a wide lead in a wide-narrow or wide-narrow-wide configuration, the potential drop and electrical field is concentrated at the point of connection. This is \eg seen in calculations of a graphene nano-ribbon connected to graphene\cite{Papior2016,Leitherer2019}. The voltage-drop dipole and resulting change in charge distribution leads to current-induced forces which can be related to the change in bond-charges\cite{Brandbyge2003,Leitherer2019}. However, it is clear that although the voltage-drop and associated electrical field is localized at the scatterer, the current is present throughout the system. This leads to forces and strains entirely related to the local current density since the local field is vanishing\cite{Leitherer2019}. 

\begin{figure*}[!bth]
	\includegraphics[width=0.95\linewidth]{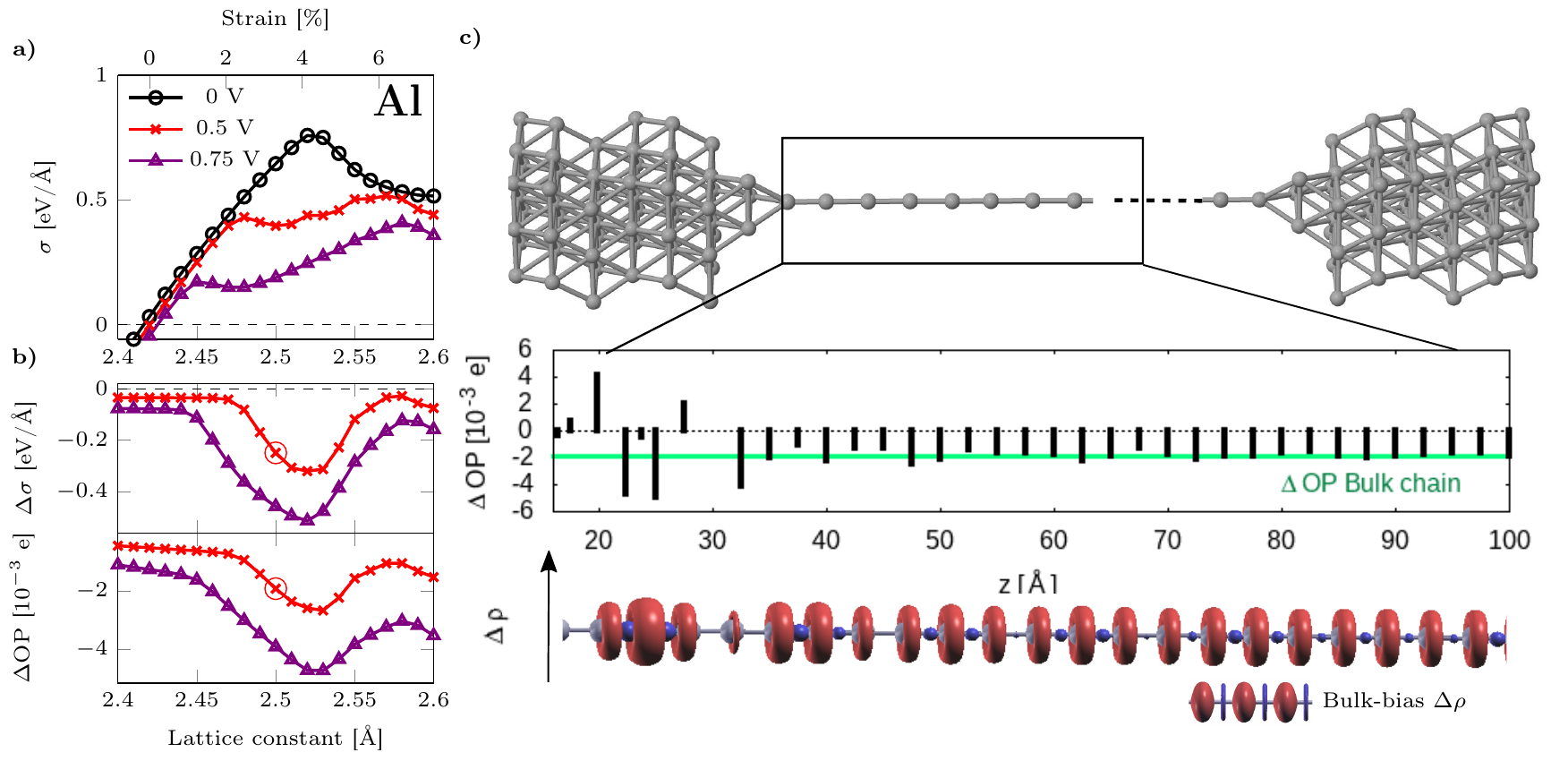}
	\caption{(a) Stress (force) as a function of strain (lattice constant) of 1D Al bulk chain for different values of the bulk-bias voltage (b) Change in stress with bias, $\Delta\sigma(V) = \sigma(V)-\sigma(0)$, (top) and change in overlap population (OP) (bottom) in the Al  bulk chain (c) DFT-NEGF calculation of 1D Al chain with bond length $\mathrm L=2.5\text\AA$ ~ connected to 3D Al electrodes. Below: Bias induced OP at $0.5\,\mathrm V$: Black bars represent $\Delta$OP(z) from DFT-NEGF, the green line $\Delta$OP from the bulk-bias calculation. Bottom: Induced charge density along the chain with comparison of the bulk-bias calculation (insert). The induced charge density and OP converge in the chain far away from the electrode interface ($z>75$ \AA), where the electrostatic potential is constant. }
	\label{fig:comp}
\end{figure*}

Thus, it is interesting to consider the role of the current {\em alone} and the related charge redistribution separate from the voltage-drop. Here we propose a very simple scheme based on standard DFT with periodic boundary conditions, to calculate the effect of current on the bonding, \ie the stress-strain relation and yield-strength, in the presence of a strong current.
To this end we use the ideal, ballistic distribution function which depends on the group velocity, and fill the Bloch states according to their band velocity projected along the applied external electrical field, $\hat{\bf e}$. Thus, we consider the same current-density distribution in all unit-cells. This nonequilibrium distribution will shift and deform DFT bandstructure, $\varepsilon_{\bfk,i}$, where $i$ is band index, compared to the equilibrium case. The basic idea is sketched in \Figref{fig:1Dexample} for a simple one-dimensional model bandstructure.

In the following $V$ and $\Vdir$ denotes the magnitude of the applied bias, and the field direction unit vector, respectively, while
\begin{align}
\label{eq:velocity}
  {\bfv}_{\bfk,i}&=\frac1\hbar\frac{\partial\varepsilon_{\bfk,i}}{\partial \bfk},
  \\
  \label{eq:velocity:projection}
  p_{\bfk,i} &= \Vdir\cdot\bfv_{\bfk,i},
\end{align}
with $\bfv_{\bfk,i}$ being the band velocity of band index $i$ and $p_{\bfk,i}$ the velocity projected in the field direction. We will in the following denote the bias, $V$, applied in this way as a bulk-bias.
We will define ``left'' and ``right'' moving states according to the projection along $\hat{\bf e}$ and fix the chemical potentials for left and right-movers relative to a quasi-Fermi level as $\mu_L=E_F-eV/2$ and $\mu_R=E_F+eV/2$ with $V$ being the applied voltage, and the field $\Vdir$ is directed from left to right. The quasi-Fermi level, $E_F$, is determined in the DFT self-consistent cycle (SCF) such that the unit-cell is charge-neutral, as in standard DFT calculations. In practice, $E_F$ is determined in the DFT SCF cycle by using eigenvalues shifted according to their projected velocity direction,
\begin{equation}
   \varepsilon'_{\bfk,i} = \varepsilon_{\bfk,i} - \frac{eV}{2}[1 - 2\Theta(p_{\bfk,i})]\,,   
\end{equation}
with $\Theta(x)$ being the Heaviside function. States are filled according to the Fermi distribution, $n_F(\varepsilon'_{\bfk,i}-E_F)$. However, note that this shift is only applied when determining $E_F$, while the un-shifted eigenvalues are used in the calculation of total energy etc. as in usual DFT calculations. With this approach the effective change in distribution function relative to quasi-equilibrium is,
\begin{equation}
    \begin{split}
    \delta f(\bfk,i)&=\Theta(p_{\bfk,i}) \left[n_F(\varepsilon_{\bfk,i}-\mu_L)-n_F(\varepsilon_{\bfk,i}-E_F)\right]\\
  &+\Theta(-p_{\bfk,i}) \left[n_F(\varepsilon_{\bfk,i}-\mu_R)-n_F(\varepsilon_{\bfk,i}-E_F)\right]    
    \end{split}
\end{equation}
We note that $E_F$ will in general depend on the applied bias. In the following, we will consider low temperature.

\begin{figure}[!bth]
	\centering
	\includegraphics[width=0.95\linewidth]{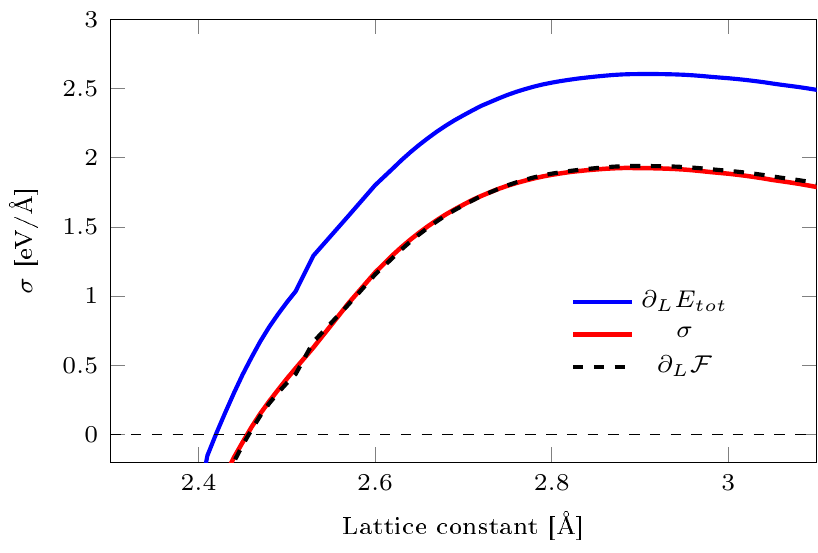}
	\caption{Stress ($\sigma$) over lattice constant of a Pt chain at 1 V bulk-bias. The red curve is the stress obtained from the Hellmann-Feynman theorem. The derivative  of  the total energy with respect to the 1D lattice constant (unit-cell size) $L$  (blue) does not include the contributions from left/right moving states $N_L$ and $N_R$, which are contained in the free energy derivative (black).}
	\label{fig:free-energy}
\end{figure}

Once the self-consistent Hamiltonian with applied bulk-bias has been calculated, we can obtain the current flowing in the structure in the direction of $\Vdir$ via,
\begin{equation}
I(V) =2 e\sum_i\!\int\!\, p_{\bfk,i}\,\delta f(\bfk,i)\,\frac{\mathrm d\bfk}{\Omega}\,,
\label{eq:IV}
\end{equation}
where $\Omega$ is the inverse Brillouin zone volume depending on the dimensionality of the system. The factor $2$ is spin degeneracy. We note that for 2D or 3D systems, we may in general obtain a current density distribution in other directions than $\hat{\bf e}$ by considering other velocity projections in \Eqref{eq:IV}. However, in the remaining of the paper, we will consider 1D systems. In this case, at zero temperature, we can rewrite \eqref{eq:IV} as,
\begin{equation}
    I(V) = \frac{2 e}{h} \!\int_{\mu_R}^{\mu_L} \!\! N(\varepsilon)\,  d\varepsilon\,,
\end{equation}
where $N(\varepsilon)$ denotes the number of bands crossing the energy $\varepsilon$. For a single band in the entire voltage window, $[\mu_R;\mu_L]$, we get,
\begin{equation}
I=\frac{2 e^2}{h} V = G_0\, V\approx 77.5\, \mu\mathrm{A}/\mathrm{V}\,.
\end{equation}

The stress (force for 1D) is calculated using the Hellmann-Feynman theorem\cite{Nielsen1985}. For the periodic systems we may evaluate the total energy per unit-cell, $E_{\mathrm{tot}}$. Importantly, we note that the derivative of this with respect to unit-cell length does not correspond to the stress in the case of finite voltage/current. We should instead consider the nonequilibrium free energy\cite{Sutton2004}, $\mathcal F$, and include the chemical potentials of left and right moving states,
\begin{equation}
	\mathcal{F} = E_{\mathrm{tot}} - \mu_L N_L - \mu_R N_R\,,
\end{equation}
where $N_L$/$N_R$ is the number of left/right moving states (according to $\Vdir$) in the unit-cell. The importance of the nonequilibrium contribution to $\mathcal F$ and the force is illustrated by an example in Fig.~\ref{fig:free-energy}.

\section{Results}

\subsection{DFT-NEGF and bulk-bias comparison}

We first apply the bulk-bias to a 1D Al chain to illustrate the method, and compare it to the same Al chain connected to 3D Al electrodes in a transport calculation at finite bias using DFT-NEGF\cite{Brandbyge2002}. In Fig.~\ref{fig:comp}(a), the stress ($\sigma$) over the lattice constant of the 1D Al chain at different bulk-bias voltages is shown. The change in lattice constant defines the strain $\varepsilon=(L-L_0)/L_0$, where $L$($L_0$) is the (equilibrium) lattice constant. For all bias points, the stress in the Al chain rises linearly to a maximum, after which it becomes strongly nonlinear (a more detailed discussion and comparison to other materials is given below). We compare the change in stress with bias, $\Delta\sigma(V) = \sigma(V)-\sigma(0)$, to the change in bond-population in Fig.~\ref{fig:comp}(b). This illustrates the relation between the current-induced stress due to the change in bond-charge in agreement with earlier studies\cite{Brandbyge2003,Brand2019}.

Figure~\ref{fig:comp}(c) depicts the results of a DFT-NEGF calculation where a long 1D Al chain is connected to 3D bulk Al electrodes (top panel). The potential drop and electrical field in this system is concentrated at the point of connection close to the higher chemical potential. That means that inside the chain, sufficiently far away from the electrode interface, there is no influence of the voltage drop or field, and the induced charges and resulting strain in the chain, are entirely due to the local current density. 
The charges and strains in this region resemble those from the bulk-bias calculation. This is demonstrated in the bottom panels of Fig.~\ref{fig:comp}(c) where we consider the change in electrons residing in the bonds, \ie overlap  population (OP), and the overall charge density along the junction. Inside the chain the change in overlap population is $\Delta OP\sim -2\,10^{-3} e$, in good agreement with the infinite chain results (\Figref{fig:comp}b lower panel, circle). The real-space change in density, $\Delta\rho$, is further compared in the lower panel in Fig.~\ref{fig:comp}(c) and the lower right inset.
In general, a discrepancy can be attributed to the difference in the actual current distribution (coming from the 3D electrodes) vs. the bulk-bias distribution, which is based on bulk bands of the 1D chain, where the reflection at the 3D-1D interface is neglected.

\subsection{Bands, DOS, COOP and overlap population}

\begin{figure}[tbh]
	\centering
	\includegraphics[width=0.8\linewidth]{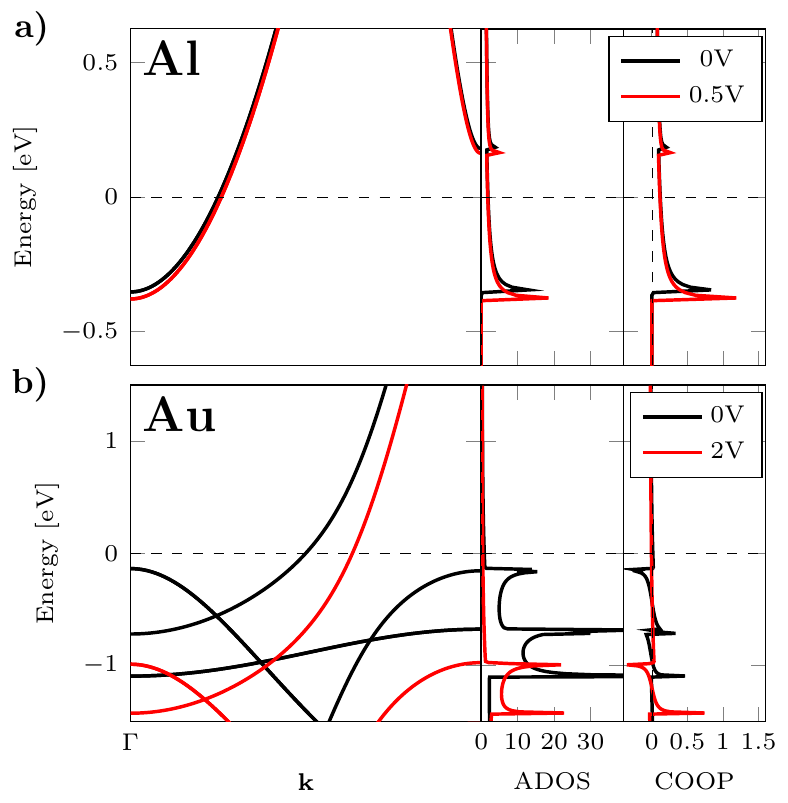}
	\caption{ Band structure, density of states(DOS), and crystal orbital overlap population(COOP) of (a) bulk Al chain with bond length $L=2.5\,\text\AA$ at $0\, \mathrm V$ (black) and $0.5\, \mathrm V$ (red), (b) bulk Au chain with bond length $L=2.72\,\text\AA$ at $0\, \mathrm V$ (black) and $2.0\, \mathrm V$ (red) }
	\label{fig:bands}
\end{figure}

\begin{figure*}[tbh]
	\includegraphics[width=0.95\linewidth]{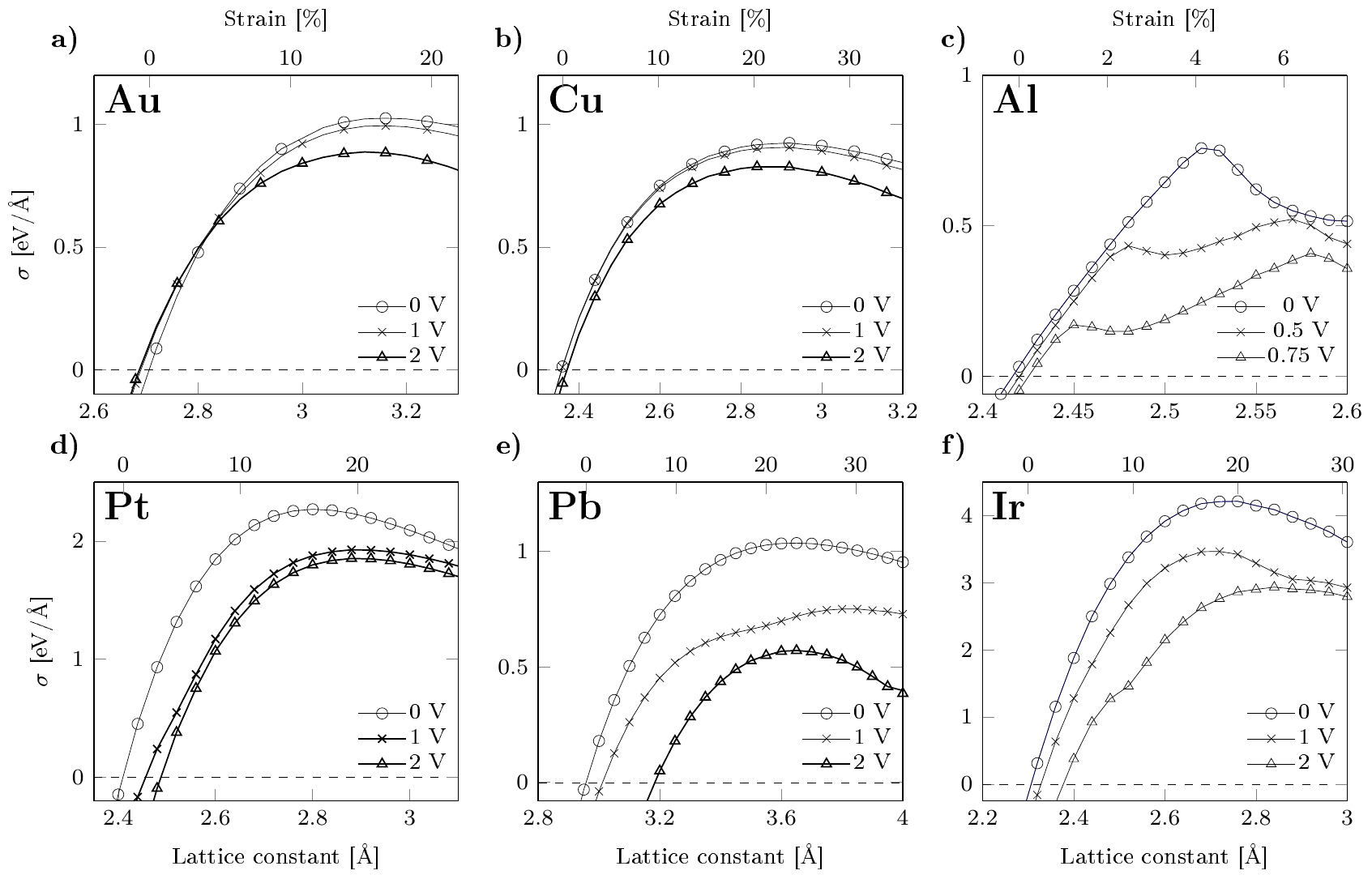}
	\caption{Stress (force) as a function of strain (lattice constant) for one-dimensional, atomic chains at different bulk-bias voltages.}
	\label{fig:stress-strain}
\end{figure*}

We can relate the change in charge residing in the bond, \ie the overlap population (OP), with bulk-bias to the underlying electronic bandstructure.
The change in OP depends on the change in the filling of the states near the Fermi energy due to the applied bias.
In particular, filling (depleting) bonding states increases (decreases) the OP, and vice versa for antibonding states.
Fig.~\ref{fig:bands} (a) demonstrates this principle for the Al bulk chain. The band structure, density of states (DOS), and crystal orbital overlap population (COOP) at $0\,\mathrm V$ (black) and $0.5\,\mathrm V$ (red) near $E_F$ are shown. The bands are only slightly shifted by the bulk-bias, while the DOS and COOP are nearly unchanged. The positive sign of the COOP indicates that all states in the voltage window are of bonding nature. The applied bias leads to the occupation of bonding states above $E_F$, and the depletion of bonding states below $E_F$. As the amount of bonding states getting depleted is higher than the amount getting filled, the bond charge (area below COOP) decreases in comparison to equilibrium. This results in bond weakening stresses in the chains.
In contrast, for the Au bulk chain (Fig.~\ref{fig:bands} (b)), we find a strong shift of the bands at $2\,\mathrm V$. The DOS is pinned to the lower chemical potential, leading to only minor changes in the occupation. This pinning is controlled by the occupation of the filled $d$-states. The small bond strengthening we see at $2\,\mathrm V$ results from the depletion of antibonding states (negative COOP). 

\subsection{Mechanical properties at nonequilibrium of 1D chains}
\label{sec:discuss}

We will now apply the bulk-bias method on 1D atomic metal chain systems, to compare the trend in the current-induced bond-weakening over the different metals. The stress in the 1D chain, which for 1D is a force ($F$), is related to the strain $\epsilon$ in terms of the linear and nonlinear elastic moduli, $E$ and $D$, respectively,
\begin{equation}
F = E\epsilon + D\epsilon^2
\end{equation}
with $D<0$, so the term will decrease the stiffness at large tensile strain. The strength or maximum tensile force corresponds to $\partial F/\partial\epsilon=0$, $F_{\mathrm{max}}=-E/2D$.  
After this point, plastic deformation occurs.

The stress-strain curves of 1D chains of different metals at finite bulk-bias voltages are shown in Fig.~\ref{fig:stress-strain}.
The overall behaviors are remarkably different. 
However, for all materials presented, we find that the stress decreases with bias. Further, except for Au, we find an increase of the equilibrium lattice constant (dashed line for zero stress) \ie the chains expand with bias, corresponding to a weakening of the bond strength. 
Au and Cu Fig.~\ref{fig:stress-strain}(a,b) exhibit a relatively weak dependence on the applied bulk-bias while the lattice constant show a minor decrease for Au in contrast to Cu, where the lattice constant increases by $1\%$ at $2\,\mathrm V$. This behavior is in strong contrast to the case of Al (Fig.~\ref{fig:stress-strain}c), which is very sensitive to both applied strain and the applied bulk-bias.
The Al chain already becomes unstable at bias voltages below $1\,\mathrm V$. The yield point of Al is significantly lowered for a bulk-bias of $0.5\,\mathrm{V}$, while the equilibrium lattice constant remains nearly the same.  The metals Pt, Pb and Ir (Fig.~\ref{fig:stress-strain}(d-f)) are also significantly influenced by the bulk-bias compared to Au and Cu, in both their equilibrium bond length and stress maximum. However, Pt and Ir are able to sustain a higher force than Au even at $2\,\mathrm V$. 

\begin{figure}[!h]
	\centering
	\includegraphics[width=0.95\linewidth]{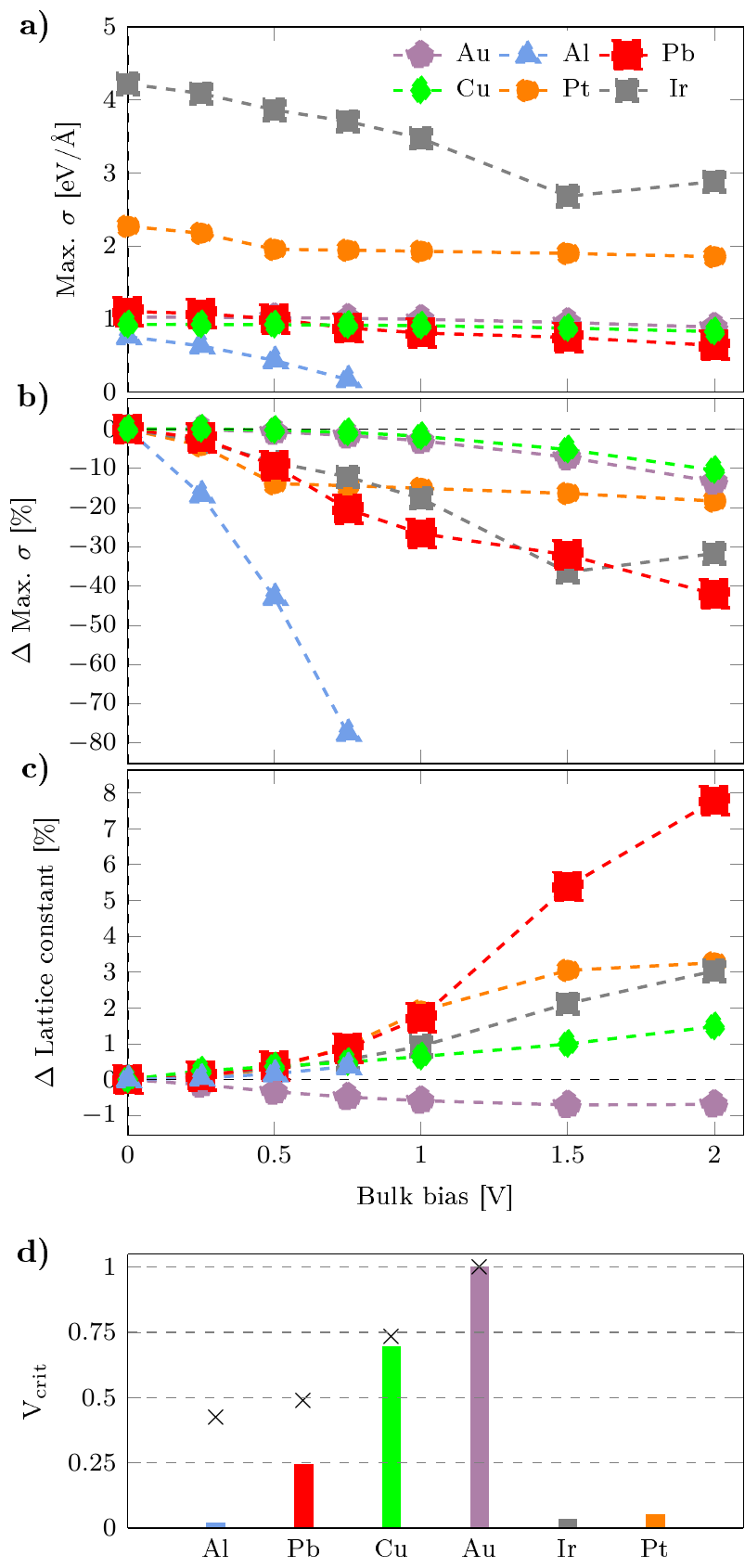}
	\caption{ (a) Ultimate stress over bulk-bias of the 1D chains shown in Fig. \ref{fig:stress-strain}. (b) Change of lattice constant with bias voltage and (b) change of maximum stress in $\%$. (d) Calculated critical voltages (normalized to V$_{\mathrm{crit}}$ of Au).  The black crosses are the experimental switching voltages from Ref.~\citenum{Ring2020}.
	}
	\label{fig:compare}
\end{figure}

In order to attempt a simple comparison with the experimental data we compare the maximum stress, the change in stress, and the change in lattice constant between the metals in Fig.~\ref{fig:compare} (a-c). First, we note that the sequence in maximum sustainable stress,  $\sigma_{\mathrm{max}}$,  for the metals follow a sequence which is not changed by the bulk-bias up to $2\,\mathrm V$. Only the case of Al yield an unstable negative stress above $\sim 0.75\,\mathrm V$. Atomic contacts down to a single atom width of Au, Cu, Pb and Al were studied in mechanically controlled break junction experiments by Ring \etal\cite{Ring2020} where characteristic threshold voltages corresponding to changes in the atomic structure were extracted as a function of contact size. These threshold voltages have shown to follow (in decreasing order) the material sequence (Au, Cu, Pb, Al) for contact conductances 1--6 $G_0$. Interestingly, in our calculation, we find the same sequence in the maximum stress, the induced stress, and in the change in lattice constant (Fig. \ref{fig:compare} (a-c)). 

We may as a rough, simple measure define a characteristic critical voltage from our calculation as $V_{\mathrm{crit}} = W_0/(dW/dV)$, where $W$ is the work needed to break the chain obtained by integrating the stress from $0$ to $\sigma_\mathrm{max}$, using as $dW/dV$ the low-bias slope, and $W_0$ the equilibrium work. 
In Fig.~\ref{fig:compare}d we plot as black crosses the experimental switching voltages from Ref.~\citenum{Ring2020} for the conductance corresponding to the infinite chains at zero voltage (Au and Cu: 1G$_0$, Pb: 3 G$_0$, Al: 2 G$_0$). For comparison, the absolute values are normalized to the critical voltage of Au. Most notably, our simulations reproduce the sequence of critical voltages observed experimentally. These experimental findings are furthermore in accordance with earlier observations showing how Au single atom wide contacts\cite{Nielsen2002,Smit2004} can withstand voltage bursts beyond $2\,\mathrm V$, while for Al\cite{Mizobata2003} this is below $0.8\,\mathrm V$, and Pt\cite{Nielsen2002} below $0.6\,\mathrm V$. It should of course be noted that while Au, Pt, and Ir are known to form chains in experiments\cite{Smit-AuPtIr-2003} in agreement with DFT\cite{Fernandez-Seivane2007}, it is not clear how well this model describes the smallest contacts of the metals.

\section{Conclusions}
We have presented a simple first principles method to estimate the role of current-induced bond weakening in ballistic atomic conductors. The method includes the role of the electronic current on the bonds and is implemented in a standard DFT code.   It is important to realize that our simple method relies on the fact that the resistivity dipoles, in principle, can be located far from the narrowest part. We may thus consider the current, present throughout the structure, and voltage drop/field separately. 
We have applied it to one-dimensional systems, but it is generally applicable for bulk periodic atomic structures in 2D and 3D as well.

In the application of the method, we have concentrated on one-dimensional atomic metal chains as a well studied benchmark system. It has been demonstrated in experiments\cite{scheer2013} how few-atom structures can be formed based on the  ``switching'' between different conductance levels due to changes in atomic rearrangements among the few atoms in the cross-section. Our prediction of metal stability against applied voltage/current is in accordance with the recent experiments\cite{Ring2020}. Clearly, many effects will play a role in this complicated rearrangement process besides the nonequilibrium bond weakening effects addressed above, such as electron-phonon coupling (Joule heating) and heat conduction, energy non-conserving forces, mechanical properties of the connection to bulk, atomic diffusion, etc. Intriguingly, first principles calculations\cite{Simbeck2012} of the electron-phonon coupling single atom chains showed that Au (and Cu) have significantly stronger e-ph coupling compared to Al, suggesting that the phonon effects (Joule heating and wind-force effects) would be less severe for Al at the atomic scale. Further, extensive first principles calculations on realistic structures and including the coupling of current to phonons\cite{Ring2020} (Joule and wind-force), but neglecting the current-induced bond weakening, were not able to reproduce the material stability sequence and found that Al was highest and Cu lowest in switching voltage for the smallest contacts and up to a conductance of 6 $G_0$.

Our results indicate that the nonequilibrium bond weakening play a central role in the effect. The change in stability with nonequilibrium may have useful applications for future atom-scale memristive devices\cite{Terabe2005,Schirm2013,Torok2020}.

\appendix
\subsection{Implementation and parameters}
We have implemented the method in the \siesta\ DFT\cite{Soler2002,Garcia2020} code, which employ a LCAO basis set.
In the LCAO basis we can readily calculate the diagonal velocity matrix element:
\begin{equation}
	\label{eq:velocity:LCAO}
	\bfv_{\bfk, i} = \frac1\hbar\left\langle \psi_{\bfk,i} \left|
	\frac{\partial {\mathbf H_\bfk}}{\partial\bfk}
	-
	\varepsilon_{\bfk,i}\frac{\partial {\mathbf S_\bfk}}{\partial\bfk}
	\right| \psi_{\bfk,i}  \right\rangle,
\end{equation}
where $\mathbf H$ and $\mathbf S$ are Hamiltonian and overlap matrices in k-space, respectively. 
The derivatives can be done analytically within LCAO using the real-space matrix elements, e.g.,
\begin{equation}
	\label{eq:velocity2:LCAO}
	\frac{\partial {\mathbf H_\bfk}}{\partial\bfk}=\sum_{\bfR} {i\bfR}\, e^{i\bfk\cdot\bfR}
	\left(
	\langle 
	\bfR|{\mathbf H}|\mathbf 0\rangle-\langle 
	\bf0|{\mathbf H}|\mathbf R\rangle
	\right)\,,
\end{equation}
where $\mathbf R$ denotes lattice vectors.
Note that \eqref{eq:velocity} and \eqref{eq:velocity:LCAO} are equivalent while the latter is exact regardless of the Brillouin zone sampling, contrary to the former for discretized differentation. In the case of degenerate eigenstates a decoupling based on the eigenvectors of the degenerate subspace (\emph{bra} using index $i$ and \emph{ket} using index $j$) of the velocity matrix as given by Eq.~\eqref{eq:velocity:LCAO}.

\subsection{DFT parameters}

The calculations were done using the \siesta\ code with the PBE-GGA functional for exchange-correlation and DZP basis-set. We have disregarded magnetic effects\cite{Kumar2013}.
In \siesta\ we use an optimized k-point sampling according to the bias window. 
In the bulk calculations 1000 k-points are used. 
Furthermore, we have verified our \siesta\ calculations by simulating the 1D chains using GPAW using the plane-wave basis set\cite{GPAW-2010}.
The stress-strain curves at 0V shown in this work agree very well with those obtained using GPAW.



\section{Acknowledgements}
Funding by Villum Fonden (Grant No. 00013340) and the Danish Research Foundation (Project DNRF103) for the Center for Nanostructured Graphene (CNG) is acknowledged. Computer infrastructure resources provided by DCC\cite{DTU_DCC_resource}.


\bibliography{bulkbias}

\end{document}